# Additives Influence the Phase Behavior of Calcium Carbonate Solution by a Cooperative Ion-association Process

Zhaoyong Zou,[1] Iryna Polishchuk,[2] Luca Bertinetti,[1] Boaz Pokroy,[2] Yael Politi,[1] Peter Fratzl[1*] and Wouter J.E.M. Habraken[1]

[1)] Max Planck Institute of Colloids and Interfaces, Department of Biomaterials, Potsdam, Germany

[2)] Department of Materials Science and Engineering and the Russell Berrie Nanotechnology Institute, Technion-Israel Institute of Technology, Haifa, Israel

**ABSTRACT** Amorphous calcium carbonate (ACC) has been widely found in biomineralization, both as a transient precursor and a stable phase, but how organisms accurately control its formation and crystallization pathway remains unclear. Here, we aim to illuminate the role of biologically relevant additives on the phase behaviour of calcium carbonate solution by investigating their effects on the formation of ACC. Results show that divalent cations like magnesium ($Mg^{2+}$) ions and negatively charged small organic molecules like aspartic acid (Asp) have little/no effect on ACC formation. However, the particle size of ACC is significantly reduced by poly(Aspartic acid) (pAsp) with long chain-length, but no effect on the position of the phase boundary for ACC formation was observed. Phosphate ($PO_4^{3-}$) ions are even more effective in reducing ACC particle size, and shift the phase boundary for ACC formation to lower concentrations. These phenomena can be explained by a cooperative ion-association process where the formation of ACC is only influenced by additives those are able to attract either $Ca^{2+}$ ions or $CO_3^{2-}$ ions and, more importantly, introduce an additional long range interaction between the $CaCO_3^0$ complexes and promote the phase separation process. The findings corroborate with our proposed model of ACC formation via spinodal decomposition and provide a more realistic representation of how biology can direct mineralization processes.

## INTRODUCTION

During the formation of crystalline biominerals, transient amorphous phases like amorphous calcium phosphate (ACP) and amorphous calcium carbonate (ACC) are likely present as precursors, and interact with a pre-deposited organic matrix.[1] Soluble acidic macromolecules and some inorganic ions in the matrix are known to be effective in controlling the nucleation and growth of crystalline phases and stabilizing amorphous precursors.[1b, 2] However, little is known about their precise role in controlling the formation of these amorphous phases. One of the main reasons is that even for the system without additives the formation mechanism of amorphous phases from supersaturated solution has not been well-understood. Without this knowledge, there is no good way of evaluating the effect of additives on the formation of these minerals. This has changed only recently, where the formation of ACP at physiological conditions has been demonstrated to proceed via the aggregation of soluble ion-association complexes of calcium and phosphate ions, a process that follows nucleation theory.[3] The precipitation of ACC, however, seems to proceed via liquid–liquid phase separation, where no energetic barrier needs to be overcome as required for the nucleation process and any slight deviation from the mean concentration will spontaneously amplify. This mechanism was proposed based on many experimental indications.[4] For example, a polymer-stabilized liquid calcium carbonate phase was found to precipitate in a 20 mM $CaCl_2$ solution with high concentration of poly(Aspartic acid) (pAsp), which was initiated by the slow diffusion of ammonium carbonate vapour.[4a] An emulsion-like precipitate was also observed after mixing 0.01 M solutions of $CaCl_2$ and $Na_2CO_3$, which developed into ACC spheres via densification.[4b] Furthermore, recent molecular dynamics simulations predicted the formation of a dense liquid phase through liquid–liquid separation in supersaturated calcium carbonate solution.[5]

According to the spinodal decomposition theory, the fastest growing wavelength of concentration fluctuation will determine the size of the domain and this characteristic length scale diverges at the spinodal line. Therefore, in our previous study,[6] by evaluating the ACC particle size dependency on calcium carbonate concentration (from 5 mM to 40 mM) and temperature (from 4 °C to 48 °C), we have determined the phase boundary for ACC formation in pure calcium carbonate solutions to be 3 – 4 mM, where the inverse of ACC particle size extrapolates to zero. This opens a unique opportunity to further understand the influence of additives on ACC formation by studying their effects on the position of the phase boundary (spinodal line). This can tell us whether additive-induced changes in shape and particle size, which are commonly observed, are a result of changes in the correlation length of the unstable solution (which changes the position of the spinodal line) or due to kinetic factors like changing surface energy or the spatial distribution of solute species. Gower et al.[4a] showed that negatively charged pAsp with an average molecular weight (Mw) of 8600 is able to induce the formation of a so-called polymer-induced liquid precursor (PILP) phase of calcium carbonate, which consists of shapeless droplets of large dimensions instead of the typical nanospheres. This behavior suggests that the droplets are formed near the spinodal line where dimensions diverge to infinite. Recently, by using a high concentration of polystyrene sulphonate (PSS) with an average Mw of ~70000, Smeets et al.[7] showed that the binding of calcium ions with PSS to form Ca–PSS globules promoted the formation of ACC by locally increasing the supersaturation level and inhibiting the crystallization of crystalline calcium carbonate phases. In addition, previous studies showed that the particle size of ACC can be significantly decreased by just adding a small amount of additives equally dispersed into the solution, such as pAsp,[8] PSS,[9] or poly(acrylic acid) (PAA).[9b] All these studies use highly negatively charged polymer with long chain-length, which have strong electrostatic interactions with calcium ions ($Ca^{2+}$). However, it is still not clear whether these additive-induced ACC formation is only a result of high binding affinity of these additives to $Ca^{2+}$ or the polymer chain-length is also of critical importance in these processes.

Therefore, to understand the mechanism by which additives influence the phase behaviour of calcium carbonate solution and also the formation pathway of ACC, here we systematically investigated the influence of biologically relevant additives on ACC particle size and how they incorporate in the solid phase. Next to the negatively charged polymer like pAsp, we also studied inorganic ions like magnesium ions ($Mg^{2+}$) and phosphate ions ($PO_4^{3-}$), which were known to stabilize ACC in *in-vitro* experiments. In addition, we chose to study the effect of pAsp chain-length in particular as it bridged part of the gap between ionic additives and polymers/polypeptides. The comparison to proteins is however limited as pAsp is largely unordered and lacking a defined secondary structure. Therefore, Asp monomer and pAsp with a chain-length of 10 (pAsp10), 100 (pAsp100) and 200 (pAsp200) were investigated. By comparing the effects of different types of additives, we found that, in addition to a strong electrostatic interaction with calcium ions ($Ca^{2+}$) and/or carbonate ions ($CO_3^{2-}$), the additives must also be able to introduce long range interactions between the calcium carbonate complexes ions during phase separation in order to have an effect on the formation of ACC.

## RESULTS and DISCUSSION

Using the procedure described in the previous study,[10] the influence of Asp and pAsp with different chain-lengths on the formation of ACC was investigated. Briefly, ACC was synthesized by dosing 0.25 – 2 mL of 1 M $CaCl_2$ solution into 48 – 49.75 mL of $Na_2CO_3$ solution (5 – 40 mM) with different amounts of these additives where the amino acid to calcium molar ratio ranged from 0.3% to 6%. In the pure system without any additives, the particle size of ACC was found to increase with decreasing initial concentration of calcium carbonate.[6] As shown in Figure 1a, 1.5% of Asp had no influence on the particle size of ACC at all concentrations, while 0.3% and 1.5% of pAsp200 significantly decreased the particle size of ACC. For example, at a concentration of 5 mM, the average diameter of ACC spheres decreased from ~200 nm for the pure system to ~90 nm in the presence of only 0.3% of pAsp200. The decreasing ACC particle size suggests that the solution become more unstable in the presence of pAsp200 and the critical concentration might be shifted to lower calcium carbonate concentrations. However, with increasing initial carbonate concentrations, the effect of pAsp200 on ACC particle size became less pronounced. This could be due to the factor that the solution is more unstable at higher calcium carbonate concentrations and pAsp is not able to further destabilize the solution.[6] Because of this, the data could not be fitted using the same equation as for the pure system to extrapolate the critical concentration at the spinodal line, which is based on the stability of the system. To test whether pAsp shifted the spinodal line or not, the pH and $Ca^{2+}$ activity of the solution were measured for the experiments at lower calcium carbonate concentrations from 2 mM to 5 mM (Figure S1), which suggests that critical concentration at the spinodal line is also at ~ 3.5 mM, similar to the pure system.[6] Therefore, the shift of the spinodal line induced by pAsp might be too small to be observed experimentally.

The effects of chain-length and charge of pAsp on ACC formation were further investigated at a low concentration of 5 mM. For Asp monomer, increasing the amount to 6% still had little effect on ACC particle size (Figure 1b). This observation is consistent with a study by Tobler et al.,[11] which showed that both Asp and glycine with an amino acid to calcium molar ratio from 20% to 670% had no effect on ACC particle size as well as its structure. However, the polymer, even at a chain-length of 10, could decrease ACC particle size significantly and the particle size decreased with increasing initial concentration of the polymer (Figure 1b). With increasing chain-length to 100 (pAsp100), the decrease in ACC particle size was even more dramatic. However, further increasing the chain-length to 200 (pAsp200) or with random chain-length (~14-80, referred to as pAspX), the effect was similar to pAsp100, suggesting that when the chain-length of pAsp reached a certain value, the difference in chain-length became less important. To test the importance of the negative charge of pAsp, we also performed experiments with a non-charged polypeptide: poly(Asparagine) (pAsn) with random chain-length (~32-97) (pAsnX). The difference between pAsn and pAsp is the side chain, which for pAsp has a negatively carboxyl (-$COO^-$) group and for pAsn has an uncharged amide (-$CONH_2$) group at the pH conditions used here (pH=11.1). Results showed that pAsnX had little effect on ACC particle size and even at 6% the average diameter only slightly decreased to ~160 nm (Figure 1b). This suggests that the negative charge is required for a polymer to influence the formation of ACC, which binds with the positively charged $Ca^{2+}$ ions.

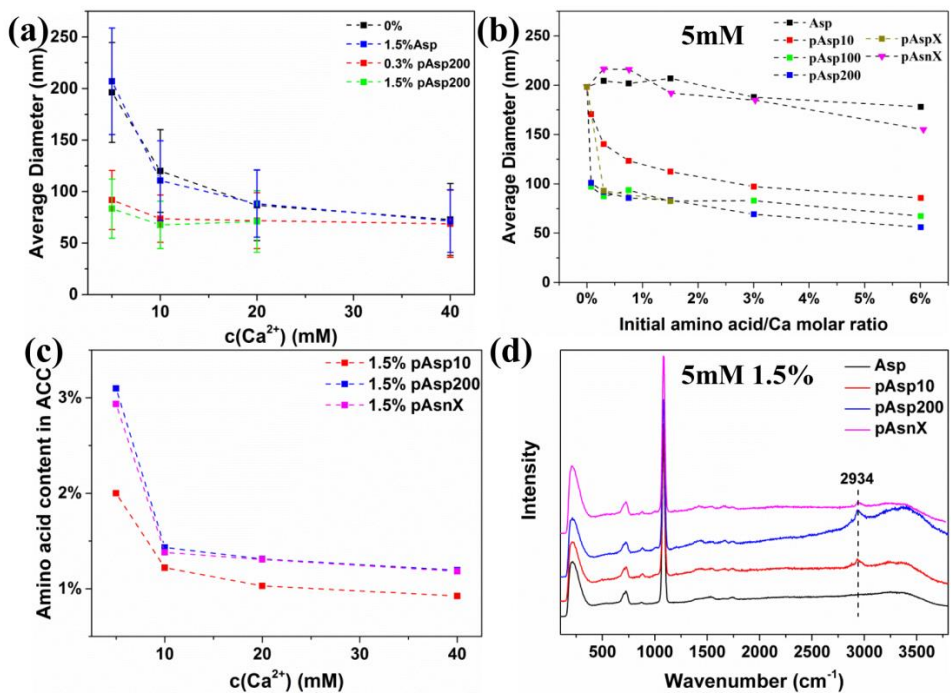

**Figure 1.** (a-b) Average diameter of ACC particles precipitated in the presence of different amount of additives and at different concentrations of $Ca^{2+}$. (c) The amount of amino acid incorporated in ACC particles precipitated at different concentrations of $Ca^{2+}$ in the presence of 1.5% of pAsp10, pAsp200 and pAsnX. (d) Raman spectra of ACC prepared at $Ca^{2+}$ concentration of 5 mM in the presence of 1.5% of additives. The dotted line refers to typical CH-vibrations of amino-acid at 2934 cm$^{-1}$.

The incorporation of amino acid in ACC was quantitatively determined by amino acid analysis (Figure 1c) and qualitatively analysed by Raman spectroscopy (Figure 1d and Figure S2). Results showed that Asp, which had no influence on ACC particle size, did not incorporated in ACC, while pAsp was incorporated in ACC and a higher amount of pAsp in ACC was observed for pAsp with longer chain-length or at lower carbonate concentrations, which were consistent with their effects on ACC particle size. However, surprisingly, pAsn was incorporated in ACC more efficiently than pAsp10, although it showed little effect on ACC particle size. The reason for this could be that pAsn do interact with either calcium or carbonate ions, which enable its incorporation in ACC, but it is not able to introduce additional effective interaction between $CaCO_3^0$ complexes. These results suggest that unlike the effect of additives on ACC particle size, a strong electrostatic interaction between the additives and $Ca^{2+}$ ions is not a necessity for the incorporation of these additives in ACC, which is mainly determined by the chain-length of the additives.

To understand these phenomena, we first have to evaluate the mechanism of phase separation of the pure system before taking into account the possible role of additives. $Ca^{2+}$ ions and $CO_3^{2-}$ ions are expected to form $CaCO_3^0$ complexes in solution, which might be the first step towards the formation of the amorphous phase.[5] In the presence of Asp or its polymer, negatively charged carboxyl groups can also bind to $Ca^{2+}$ ions and the $Ca^{2+}$ ion chelating ability increases with increasing the degree of polymerization.[12] The stability constant for the binding of $Ca^{2+}$ ions to Asp is only 40 l/mol at ionic strength of 0.1 at 25 °C,[13] which is much lower than the stability constant of $CaCO_3^0$ complexes (1700 l/mol).[14] However, the stability constant for the binding of $Ca^{2+}$ ions to pAsp with a chain-length of 72 (pAsp72) at high pH (>7) is 1047 ($10^{3.02}$) l/mol from the experimental data and 3311 ($10^{3.52}$) l/mol if assuming that the activity coefficients of different deprotonated pAsp species have the same values as those of calcium-polyaspartate complexes.[15] This value is comparable to that of $CaCO_3^0$ complexes, and therefore a significant amount of the $Ca^{2+}$ ions bind with pAsp (and not with Asp) instead of with $CO_3^{2-}$ ions. However, only strong binding of the additive with $Ca^{2+}$ ions is not enough to change the particle size. A recent study investigated the influence of citrate on the formation of ACC at an initial calcium carbonate concentration of 6.5 mM and results showed that the average diameter of ACC did not change even at a citrate/$Ca^{2+}$ molar ratio up to 10%.[16] This is surprising considering the fact that the stability constant for the complexation of $Ca^{2+}$ and citrate (~$7.0\times10^4$ l/mol) is more than 40 times higher than that of $CaCO_3^0$.[17] Compared with citrate and Asp, another important characteristic of pAsp is that being a polymer it can locally concentrate the $Ca^{2+}$ ions and the $CaCO_3^0$ complexes through the chain. According to the spinodal decomposition theory, it is the concentration fluctuations that determine the phase separation process and the final particle size of ACC.[6] Therefore, pAsp can interact with several $CaCO_3^0$ complexes and introduce an additional effective interaction between the $CaCO_3^0$ complexes, which accelerate the propagation of the concentration fluctuation of $CaCO_3^0$ and promote the phase separation process. To achieve this, the length of pAsp should be long enough and therefore Asp monomer has no effect on the phase separation process. This also explains why pAsp100, pAsp200 and pAspX showed similar effect on ACC formation, because as long as pAsp has a certain chain length to introduce such an additional long range interaction it become less important. Since the effective interaction in the pure system increases with increasing calcium carbonate concentrations, it is expected that ACC particle size would be less influenced by pAsp at higher calcium carbonate concentrations, which is also consistent with the experimental data.

Indeed other additives like poly(sodium 4-styrenesulfate) (PSS),[9] PAA,[9b] PAA-containing double hydrophilic block copolymers[18] and polycarboxylic acid made from 70 wt% acrylic acid and 30 wt% maleic acid (Mw = 70 000 g/mol)[19] also significantly decreased the particle size of ACC. Similar to pAsp, they are all large molecules containing many negatively charged groups ($SO_3^-$ group and $COO^-$ group). To test whether positively charged polymer can have similar effect or not, ACC was synthesized at a calcium carbonate concentration of 5 mM in the presence of 2% of poly(Arginine) and the average particle size was also significantly decreased to ~65 nm, which is comparable to the effect of pAsp. This suggests that positively charged polymers, which bind with $CO_3^{2-}$ ions, can also introduce an additional effective interaction between $CaCO_3^0$ and influence the formation of ACC.

As a comparison to the organic additives, we also investigated the influence of inorganic additives on the formation of ACC, including other divalent cations ($Me^{2+}$) like $Mg^{2+}$, $Sr^{2+}$ and $Ba^{2+}$ ions and anions like phosphate ions ($PO_4^{3-}$). For $Me^{2+}$ cations, our results (Figure 2a) showed that the average diameter of ACC was slightly decreased by 10% and 20% of $Mg^{2+}$ ions and 20% of $Sr^{2+}$ ions. For $Ba^{2+}$ ions, the decrease in ACC particle size was more prominent than $Mg^{2+}$ ions and $Sr^{2+}$ ions, but only at lower carbonate concentrations where the unstable solution approached the spinodal line.[6] However, although these divalent cations showed little influence on the particle size of ACC, their incorporation in ACC were starkly different. The $Mg^{2+}$ content in all ACC samples was significantly lower than the initial value in solution (Figure 1b), but it increased linearly with increasing concentration of carbonate, which is consistent with previous studies.[20] The $Sr^{2+}$ content in ACC was much higher than $Mg^{2+}$, but it increased non-linearly with increasing concentration of carbonate, reaching ~17% at 40 mM. More interestingly, the $Ba^{2+}$ content in ACC even decreased slightly from ~21% to ~20% with increasing concentration of carbonate from 5 mM to 40 mM. All together, these results showed that the ability of these cations to incorporate in ACC increased with increasing ionic radii ($Mg^{2+}<Sr^{2+}<Ba^{2+}$), but the incorporation efficiency strongly depends on the total concentration of carbonate. In agreement with our findings, a recent study showed that when $Mg^{2+}$, $Ca^{2+}$, $Sr^{2+}$ and $Ba^{2+}$ ions were coprecipitated with $CO_3^{2-}$ ions, the preferential incorporation of larger cations in the amorphous precipitates was clearly observed at lower concentration of carbonate (10 mM), but less prominent at higher concentration of carbonate (100 mM).[21] These results suggest that the relative amount of cations in the precipitate depends on not only the relative complexation ability (coulombic interaction) of the cations with carbonate ions, but also the concentration of carbonate, which determines the stability of the solution. To understand how these divalent cations change the phase behavior of the solution, the particle size data were fitted using the same equation as used in the previous study.[6] Results (Figure S3) showed that the concentrations at the spinodal line ($c_{sp}$) for $Mg^{2+}$ ions and $Sr^{2+}$ ions were similar to the pure system, but for $Ba^{2+}$ ions the value was slightly lower, suggesting that only with $Ba^{2+}$ ions, the solution was more unstable than the pure calcium carbonate solution under the same condition. This corresponds very well to the incorporation of these cations in ACC where there is a preferential incorporation of $Me^{2+}$ ions over $Ca^{2+}$ ions in the solid only in the case of $Ba^{2+}$ ions. It suggests that the formation of $BaCO_3^0$ complexes is kinetically favoured over $CaCO_3^0$ complexes under the current conditions, which will also introduce an additional effective interaction between $CaCO_3^0$ complexes and promote the phase separation of ACC with smaller particle size.

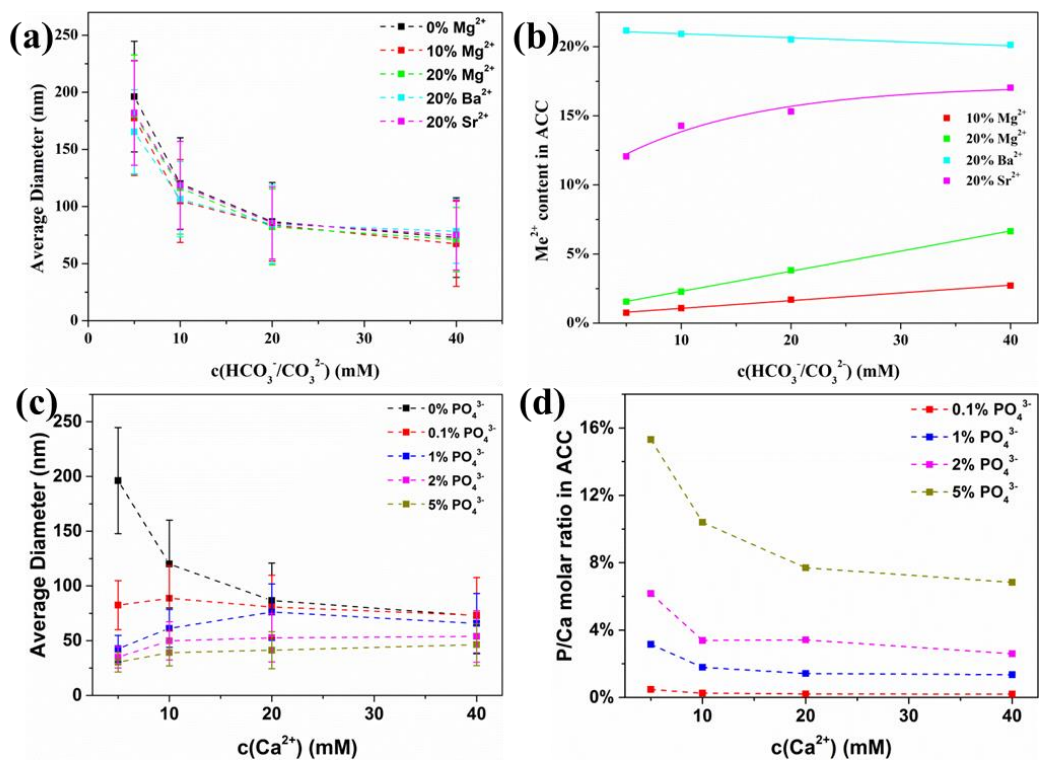

**Figure 2.** Average diameter of the ACC spheres prepared at a carbonate concentration from 5 to 40 mM in the presence of $Me^{2+}$ (a) and $PO_4^{3-}$(c), and the corresponding $Me^{2+}$ content (b) and $PO_4^{3-}$ content (d) in the ACC. The error bars indicate the standard deviation of the observed size distribution.

The influence of phosphate ions on the formation of ACC was investigated by replacing part of the $Na_2CO_3$ with the $NaH_2PO_4$, and the initial P/Ca molar ratio ranged from 0.1% to 5%. Although the pH of the reaction solution (11.1-11.4) is lower than the dissociation constant of $HPO_4^{2-}$ (pKa = 12.3), the stability constant of $CaPO_4^-$ complex ($2.9\times10^6$ l/mol) (complexation of $Ca^{2+}$ and $PO_4^{3-}$) is ~5000 times higher than that of $CaHPO_4^0$ / $Ca(HPO_4)_3^{4-}$ complexes (complexation of $Ca^{2+}$ and $HPO_4^{2-}$ in $CaHPO_4^0$ ~300 l/mol and in the $Ca(HPO_4)_3^{4-}$ complex ~750 l/mol),[3, 22] suggesting that $PO_4^{3-}$ is the dominant species to interact with $Ca^{2+}$ ions, which is confirmed by the Infrared and Raman analysis of ACC (see below). Similar to pAsp, phosphate ions also significantly reduced the average particle size of ACC, especially at lower concentrations of carbonate ions (Figure 2c). Interestingly, phosphate ions were more effective than pAsp and the average diameter even decreased with decreasing initial total concentrations. The overall phosphate content (P/Ca molar ratio) in ACC (Figure 2d) was higher than the initial phosphate concentration in solution and it increased with decreasing total carbonate concentration, suggesting that phosphate ions were preferentially incorporated in ACC as compared to the carbonate ions. By extending the range of initial P/Ca molar ratio to 0.06%-10% for the experiments at a calcium carbonate concentration of 5 mM, it was found (Figure S4a) that the average diameter of ACC spheres first decreased quickly with increasing initial P/Ca molar ratio and then gradually reached a plateau. On the other hand, the P/Ca molar ratio in ACC increased linearly with increasing initial P/Ca molar ratio in solution, with a partition coefficient of 3.0, indicating a 3 times enrichment of $PO_4^{3-}$ in the solid phase independent of the solution phosphate concentration. However, the partition coefficient decreased to 1.4 with increasing initial total concentration of calcium carbonate to 40 mM (Figure S4b). These results suggest that the phosphate ions were less effective in reducing the particle size of ACC and being incorporated in ACC when the total concentration of carbonate increased, where the solution became more unstable and the kinetics of phase separation increased.

Since phosphate ions were more effective than pAsp in reducing ACC particle size, it was expected that their effect on the spinodal line of calcium carbonate solution be prominent. Therefore, experiments were performed at lower concentrations in the presence of 2% of phosphate ions. Results showed that ACC spheres with an average diameter of ~30 nm can also be obtained at calcium carbonate concentrations of 2 mM and 3 mM (Figure S5). The P/Ca molar ratios in these ACC spheres were up to ~15% at 3 mM and ~30% at 2 mM, even though the initial P/Ca molar ratio was only 2% (Figure S5). It should be noted that without phosphate ions, ACC could not form at such low concentrations,[6] suggesting that $PO_4^{3-}$ ions indeed promote the phase separation of ACC from solution and shift the phase boundary of calcium carbonate to lower concentrations.

To understand what happens in the case of P-enriched ACC, we have to evaluate the formation kinetics and pathway of both calcium carbonate and calcium phosphate individually. In both systems, they tend to form complexes in solution.[3, 5] Habraken et al.,[3] furthermore, showed that calcium phosphate ion-association complexes readily aggregate into branched three-dimensional polymeric structures, which densify during time and transform into ACP according to a slow multi-step process. ACC, on the other hand, tends to spontaneously phase separate from solution.[5-6] As the stability constant of $CaCO_3^0$ complexes ($1.7\times10^3$ l/mol)[14] is ~2000 times lower than that of $CaPO_4^-$ complexes, $Ca^{2+}$ ions or $CaCO_3^0$ complexes would strongly interact with $PO_4^{3-}$ ions when the concentration of $PO_4^{3-}$ ions is more than 1/2000 of $CO_3^{2-}$ ions, corresponding to the observed decrease in ACC particle size at only 0.06%-0.1% of $PO_4^{3-}$ in the solution. Therefore, an explanation for the smaller particle size and phase separation of a phosphate-rich phase is that the complexation of $Ca^{2+}$ ions and $PO_4^{3-}$ ions and their polymerization/agglomoeration to form three-dimensional polymeric structures[3] locally concentrate the $CaCO_3^0$ complexes and then introduce an additional effective interaction between these complexes. This is similar to the effect of highly charged macromolecules like pAsp, however, instead of being a polymer to concentrate the $CaCO_3^0$ complexes along the chain, $PO_4^{3-}$ ions first form complexes with $Ca^{2+}$ ions and then polymerize or agglomerate together to increase the local concentration of $CaCO_3^0$ complexes. Moreover, $PO_4^{3-}$ ions can even push the stability region to lower concentrations and the $PO_4^{3-}$-dominated formation kinetics enables the precipitation of a molecularly mixed phosphate-carbonate species. To emphasize that indeed the calcium phosphate might behave like a polymer, we performed experiments in which we added equivalent amounts of condensed phosphates (pyrophosphate (PyroP) and polyphosphate, n = 100 (polyP)). Both showed similar effects on the particle size of ACC as the phosphate ions (Figure S6a), and their contents in ACC were also similar (Figure S6b). Raman spectra (Figure S6c) of the ACC samples further showed characteristic peak for the P-O vibrational mode of pyroP[23] and polyp[24] at 1042 $cm^{-1}$ and 1161 $cm^{-1}$, respectively, indicating the incorporation of pyroP and polyP in ACC without decomposing into phosphate ions.

The distribution of phosphate ions in ACC was then investigated with FTIR spectroscopy. As shown in Figure 3a, all spectra exhibit characteristic peaks of ACC, including the asymmetric stretching mode $v_3$ of $CO_3^{2-}$ at 1470 and 1408 $cm^{-1}$, the in-plane deformation $v_4$ of $CO_3^{2-}$ at 724 and 692 $cm^{-1}$, the out-of-plane deformation $v_2$ of $CO_3^{2-}$ at ~ 864 $cm^{-1}$ and the symmetric stretching mode $v_1$ of $CO_3^{2-}$ at 1075 $cm^{-1}$.[10] In addition, the asymmetric stretch $v_3$ of $PO_4^{3-}$ appears as a broad hump from 970 $cm^{-1}$ to 1160 $cm^{-1}$ without sharp splitting and the symmetric stretch $v_1$ of $PO_4^{3-}$ appears at ~ 952 $cm^{-1}$, suggesting that the $PO_4^{3-}$ ions in ACC also existed in an amorphous state,[25] which was also supported by the Raman spectra of the samples (Figure S7). The most prominent feature in the spectra was that the intensity of the peaks of $PO_4^{3-}$ increased significantly with increasing the initial P/Ca molar ratio (Figure 3a), corresponding to the increasing phosphate content in ACC. A closer examination of the spectra in the range from 800 $cm^{-1}$ to 1200 $cm^{-1}$ (Figure 3b) showed that not only the relative intensity of the peaks for $PO_4^{3-}$ and $CO_3^{2-}$ changed, but also their positions were shifted. Specifically, the $v_3$ of $PO_4^{3-}$ shifted to lower wavenumber and the $v_1$ of $CO_3^{2-}$ shifted to higher wavenumber. To quantify the difference, the the $v_2$ of $CO_3^{2-}$ was fitted with a Guass function and result showed that the the $v_2$ of $CO_3^{2-}$ gradually shifted from 864 $cm^{-1}$ to 868 $cm^{-1}$ with increasing phosphate content in ACC (Figure 3c). The shift of these bands with concentration, suggests that $CO_3^{2-}$ ions and $PO_4^{3-}$ ions are molecularly mixed with each other in the solid. A recent solid state NMR study of ACC formed in the presence of phosphate ions also shows that the $PO_4^{3-}$ ions exist in an amorphous state and are molecularly dispersed throughout the structure of ACC.[26] However, this doesn't mean that the composition of the P-enriched ACC spheres is homogenous. Patches of P-enriched regions might still exist in these particles. This may in turn provide information on the formation history of the amorphous phase. Therefore, to further investigate the distribution of $PO_4^{3-}$ ions in ACC and the chemical bonding of the atoms, X-ray photoelectron spectroscopy (XPS) analysis was performed. In contrast to ICP, which measures the composition of the sample bulk, XPS is a surface-sensitive quantitative spectroscopic technique with probing depth up to ~10 nm. Moreover, by changing the collection angle of the signal, it is possible to enhance the measurement of elements at the surface of the sample with ~63% of the information originating from the top 3 nm thick layer ($XPS_{surface}$) or enhances the measurement of elements at the bulk of the sample within 8-10 nm ($XPS_{bulk}$), which is a thicker surface compared to the bulk analysis from ICP measurement. Results (Figure 3f) showed that the P/Ca molar ratio obtained from $XPS_{bulk}$, $XPS_{surface}$ and ICP were consistent with each other for samples at $c(Ca^{2+})$ of 5 mM with 1% and 2% of $PO_4^{3-}$ and at $c(Ca^{2+})$ of 10 mM with 5% of $PO_4^{3-}$. However, for the sample at $c(Ca^{2+})$ of 5 mM with high initial $PO_4^{3-}$ content (5%) or at high $c(Ca^{2+})$ (20 mM and 40 mM), the P/Ca molar ratio on the surface ($XPS_{surface}$) was significantly lower than in the bulk (ICP), indicating a spatial variation in P content within these samples. These results showed that the distribution of $PO_4^{3-}$ in ACC not only depends on the initial concentration of $PO_4^{3-}$ ions in solution, but also the total concentration of calcium carbonate, which is consistent with their effects on ACC particle size. In addition, high resolution XPS (Figure S8) also showed a variation of the chemical environment of carbon and calcium in the presence of different amount of phosphate ions.

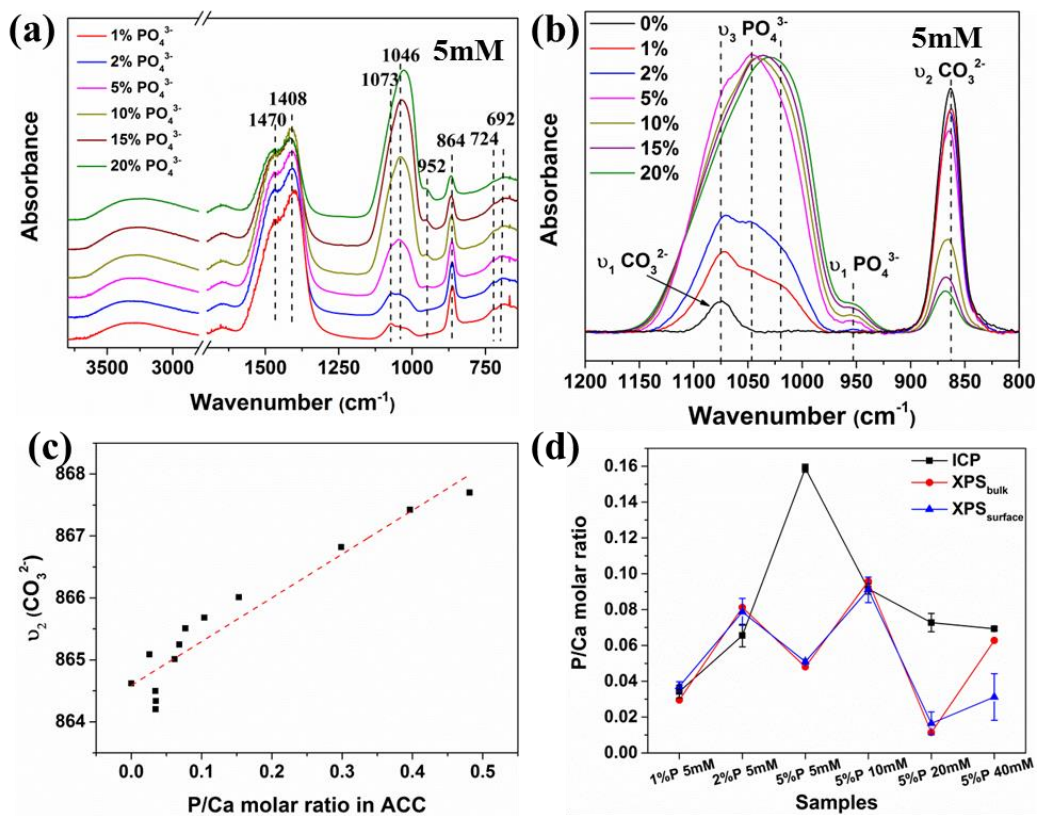

**Figure 3.** FTIR spectra of ACC spheres prepared at a carbonate concentration from 5 mM in the presence of different amount of $PO_4^{3-}$(a-b), (c) the ratio of peak area between the $v_1$ of $PO_4^{3-}$ and the $v_2$ of $CO_3^{2-}$, the ratio of the peak area between the $v_1$ of $PO_4^{3-}$ the $v_3$ of $PO_4^{3-}$ and (e) the peak position of the $v_2$ of $CO_3^{2-}$ as a function of the P/Ca molar ratio in ACC. (f) The P/Ca molar ratio measured with ICP, $XPS_{bulk}$ or $XPS_{surface}$ for ACC samples prepared at different conditions.

Based on the results of all three types of additives, mechanistic insight into the formation pathway of ACC and the effects of additives on this phase separation process can be generalized (Figure 6). In the absence of additives, $Ca^{2+}$ ions and $CO_3^{2-}$ ions first form $CaCO_3^0$ complexes, after which ACC phase separates from solution via a spinodal decomposition process where the overall concentration of $CaCO_3^0$ complexes determines the average particle size of ACC.[6] To change the particle size and promote the phase separation of ACC, the additives must be able to form complexes with $Ca^{2+}$ or $CO_3^{2-}$ ions and simultaneously be able to introduce long range interaction between the $CaCO_3^0$ complexes. Such a cooperative ion-association process can be achieved by charged polymers with long chain-length, which are able to attract many counter-ions and their associated $CaCO_3^0$ complexes along the chain. Another way to achieve this is to use ions like $PO_4^{3-}$ ions that are able to form dense structure with counter-ions. More importantly, it requires that this process is kinetically faster than the densification of $CaCO_3^0$ complexes, which depends on the concentration of both additives and calcium carbonate in solution. The change of the phase behaviour of calcium carbonate solution (i.e. position of the prospected spinodal line), however, was only clearly observed by an additive ($PO_4^{3-}$) that is able to form a less soluble salt with $Ca^{2+}$ by itself. It must be noted that as ACP was observed to form via nucleation,[3] the mechanism at which this occurs is likely in between spinodal decomposition and nucleation, and no clear spinodal or connodal line can be drawn here (also because of varying compositions).

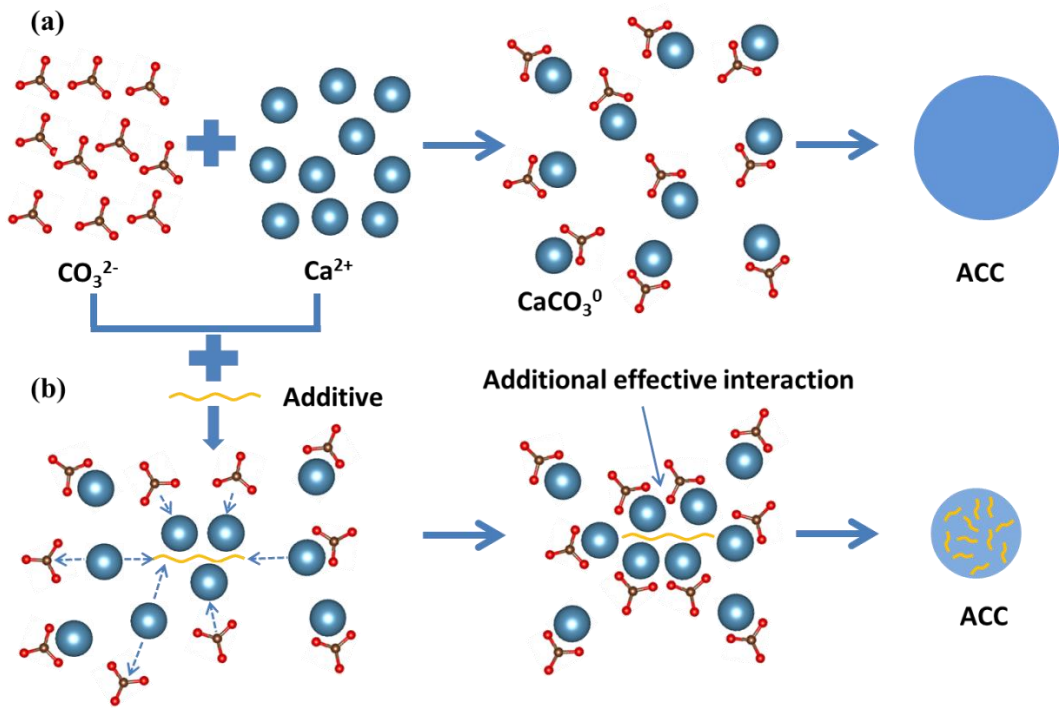

**Figure 4.** Mechanism of calcium carbonate formation: (a) In the absence of additives, the binding between positively charged $Ca^{2+}$ ions and negatively charged $CO_3^{2-}$ ions results in the formation of $CaCO_3^0$ complexes, and the effective interaction between the complexes lead to the formation of ACC when the solution is unstable. (b) In the presence of additives that are able to influence the formation of ACC, the first step is a cooperative ion-association process where the additives attract either $Ca^{2+}$ ions or $CO_3^{2-}$ ions which are associated to other counter ions. By doing so, the additives provide an additional effective interaction between the $CaCO_3^0$ complexes, which destabilizes the solution and induces the formation of ACC with smaller particle size.

As biogenic ACC biominerals also consist of spherical nanoparticles from tens of nanometer to hundreds of nanometer in diameter, the particle size could be used to understand the initial concentration of calcium carbonate where ACC is deposited, as we have speculated in our previous work.[6] However, according to the phase diagram for the pure system, an average diameter of ~30 nm, which is commonly observed in organisms, suggests an initial calcium carbonate concentration of more than 100 mM. Here, we showed that organisms might use macromolecules and inorganic ions to control the particle size of ACC even at low concentrations. In addition, by locally concentrating $Ca^{2+}$ ions or $CO_3^{2-}$ ions with these molecules, ACC can be precipitated at much lower concentrations, which also explains why mineralization of calcium carbonate can occur in undersaturated conditions. Another important observation is that additives only seem to work at low concentrations where apparently the kinetics for ACC formation is slow enough. The prospected mechanism for the additive-directed decrease in particle size assumes that there is such a time window in which the additives can interact with the ions/complexes during phase separation. Therefore, next to providing a more realistic representation of how biology can direct mineralization, this observation strengthens our theory that indeed we are reaching a phase boundary, which would be the ACC spinodal line.

## CONCLUSION

To understand the mineralization process of ACC in organisms, we systematically investigated the influence of biologically relevant additives on the phase behaviour of ACC. We have demonstrated that divalent cations like $Mg^{2+}$, $Sr^{2+}$ and $Ba^{2+}$ ions have little effect on the particle size of ACC, while their incorporation in ACC strongly depends on their dehydration kinetics. While Asp monomer shows little effect on ACC particle size, pAsp with long chain-length and anions like phosphate ($PO_4^{3-}$) ions significantly decrease the particle size of ACC and show a preferred incorporation into the ACC especially at low carbonate concentrations. In addition, $PO_4^{3-}$ ions are able to shift the phase boundary for ACC formation to lower concentrations because of their property to strongly bind and concentrate $Ca^{2+}$ ions by forming polymer-like structures during ACC phase separation. Altogether, our study suggests that additives influence the phase behaviour of calcium carbonate solution by a cooperative ion-association process where the additives are able to bind with either $Ca^{2+}$ or $CO_3^{2-}$ ions and introduce an additional effective interaction between $CaCO_3^0$ complexes during the phase separation of ACC.

## EXPERIMENTAL SECTION

**Materials and general preparative methods.** Analytical grade calcium chloride dihydrate ($CaCl_2 \cdot 2H_2O$), sodium carbonate decahydrate ($Na_2CO_3 \cdot 10H_2O$), L-Aspartic acid, poly-(α,β)-DL-aspartic acid sodium salt) with random chain-length (~14-80) (MW = 2000-11000 Da) and poly(L-asparagine) with random chain-length (~32-97) (MW = 5000-15000 Da) were purchased from Sigma-Adrich. Poly(L-aspartic acid sodium salt) with chain-length of 10, 100 and 200 were purchased from Alamanda Polymers, corresponding to a molecular weight of 1400, 14000 and 27000 Da, respectively. The 1 M calcium solution, carbonate solution with various concentrations and 1 mg/mL additive solutions were prepared by dissolving corresponding chemicals in ultrapure water. A computer controlled titration system consisting of a titration device controlling three dosing units (800 Dosino) (905 Titrando, Metrohm Ltd.) was utilized for the experiments.

**ACC synthesis in solution.** The experiments were performed at (24±1) °C in a 100 mL vessel filled with certain amount of carbonate solution with/without additives (48-49.75 mL) under stirring. 1 M calcium solution (0.25-2 mL) was dosed through the dosing unit into the reaction vessel at a rate of 10 mL per minute to ensure a rapid mixing of the two solutions and ACC precipitated immediately. The total volume of the solution after dosing were 50 mL and the concentration ratio of calcium to carbonate after mixing was 1:1. The reaction solution was monitored by using a pH electrode and a Ca ion-selective electrode. The precipitates were collected immediately after precipitation by fast vacuum filtering of the reaction solution and rinsed with ethanol. The dry powders were then stored in a vacuum desiccator for further characterization.

**Scanning electron microscopy (SEM).** Scanning electron microscopy was performed using a field emission scanning electron microscope (JEOL, JSM-7500F) working at an acceleration energy of 10 keV. Samples were not coated prior to investigation.

**Raman and infrared spectroscopy.** Raman spectra were collected using a confocal Raman microscope (α300; WITec) equipped with a CCD camera (DV401-BV; Andor), a Nikon objective (10×) and a 532 nm laser (40 accumulations, integration time 1s). Infrared spectra were recorded using a Thermo Scientific Nicolet is5 FTIR spectrometer (ATR-Diamond mode) (500 scans, resolution 2 cm$^{-1}$).

**Inductively Coupled Plasma-Optical Emission Spectrometry (ICP-OES) Analysis.** Elementary analysis was performed by ICP-OES (PerkinElmer, Optima 8000 ICP-OES Spectrometer). Sample solutions were prepared by dissolving certain amounts of as-prepared powders in diluted $HNO_3$ solution.

**Amino acid analysis.** ACC powder with known weight was hydrolyzed in 6 N HCl containing 5% phenol by vacuum sealing and heating at 110 °C for 24 h. Hydrolyzed samples were flash evaporated, resuspended in sample dilution buffer and run on a postcolumn ninhydrin-based amino acid analyzer (Sykam S433, Fürstenfeldbruck, Germany).

**X-ray Photoelectron Spectroscopy (XPS) Analysis.** Investigation of the chemical bonding of the ACC powders was performed utilizing Thermo VG Scientific Sigma Probe with monochromatic Al Kα, 1486.6 eV X-ray source, beam size 400 μm*200 μm. The samples were irradiated with monochromatic X-rays. Survey spectra were recorded with a pass energy 200 eV. The core level binding energies of the different peaks were normalized by setting the binding energy for the C1s at 285.0 eV.

## ASSOCIATED CONTENT

This information is available free of charge via the Internet.

## AUTHOR INFORMATION

**Corresponding Author**

* fratzl@mpikg.mpg.de

## ACKNOWLEDGEMENT

The authors thank Steve Weiner and Lia Addadi for useful discussions. The research was supported by a German Research Foundation grant within the framework of the Deutsch–Israelische Projektkooperation DIP. Zhaoyong Zou was supported by China Scholarship Council (CSC). B.P. acknowledges financial support by the European Research Council under the European Union's Seventh Framework Program (FP/2007-2013)/ERC Grant Agreement (no. 336077)